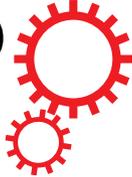

# OPEN

# Role of the particle size polydispersity in the electrical conductivity of carbon nanotube-epoxy composites

Maryam Majidian[1], Claudio Grimaldi[1], László Forró[1] & Arnaud Magrez[1,2]

Carbon nanotubes (CTNs) with large aspect-ratios are extensively used to establish electrical connectedness in polymer melts at very low CNT loadings. However, the CNT size polydispersity and the quality of the dispersion are still not fully understood factors that can substantially alter the desired characteristics of CNT nanocomposites. Here we demonstrate that the electrical conductivity of polydisperse CNT-epoxy composites with purposely-tailored distributions of the nanotube length L is a quasiuniversal function of the first moment of L. This finding challenges the current understanding that the conductivity depends upon higher moments of the CNT length. We explain the observed quasiuniversality by a combined effect between the particle size polydispersity and clustering. This mechanism can be exploited to achieve controlled tuning of the electrical transport in general CNT nanocomposites.

The relatively high level of the electrical conduction at very low filler concentrations makes carbon nanotube-polymer composites attractive materials for a wide range of applications where unaltered optical and/or mechanical properties of the host-insulating medium are required[1–5]. In these systems, as generally in other polymer nanocomposites, the electrical connections between the conducting particles are established by tunnelling of electrons across the thin polymer layer separating the conductive fillers. The enhanced electrical connectedness of polymers filled with carbon nanotubes (CNTs) is understood as being driven by the increased excluded-volume associated with the high aspect-ratios of CNTs[6,7], which has the net effect of reducing the minimal (or critical) inter-particle distance ($\delta_c$) that the electrons have to tunnel in order to establish a system-spanning tunnelling connectivity[8–14]. For this reason, polymers filled with longer CNTs are expected to conduct electricity at lower CNT concentrations, as it is generally observed in experiments[15,16]. The optimization of the electrical-mechanical performances of CNT-polymer materials is however hindered by the almost inevitable polydispersity in length ($L$) and diameter ($D$) of the nanotubes, which is thought to be one factor responsible for major discrepancies observed in the conductivities of apparently similar CNT-polymer composites[2,17].

Recently, a testable prediction about the effect of nanotube polydispersity has been made by applying the connectedness percolation theory to the liquid-state dispersions of slender, straight and polydisperse rod-like particles. It has been shown that the minimum filler loading required to establishing a system-spanning cluster of connected rods, also referred to as the percolation threshold, is inversely proportional to the weighted average $L_w = \langle L^2 \rangle / \langle L \rangle$ of the rod lengths[18–23], where the brackets denote averages over the distribution of $L$. Numerical simulations have confirmed and extended this finding by showing that the percolation threshold is a quasi-universal function of $L_w$ even for homogeneous dispersions of straight rods of intermediate aspect ratios[11,24,25].

When applied to dispersions of rods with inter-particle tunnelling, these results amount to predict that the critical tunnelling distance $\delta_c$ depends upon the rod length distribution only through $L_w$ for a given volume fraction of the nanotubes[11,13,26]. The resulting bulk conductivity $\sigma$ is thus expected to display a similar quasi-universal behaviour as a function of $L_w$, implying that knowledge of the scaled variance of $L$, $\langle L^2 \rangle/\langle L \rangle^2 - 1$, is necessary in order to control $\sigma$.

[1]Laboratory of Physics of Complex Matter, Ecole Polytechnique Fédérale de Lausanne, Station 3, CH-1015, Lausanne, Switzerland. [2]Crystal Growth Facility, Ecole Polytechnique Fédérale de Lausanne, Station 3, CH-1015, Lausanne, Switzerland. Correspondence and requests for materials should be addressed to C.G. (email: claudio.grimaldi@epfl.ch)

Although theories and simulations agree in identifying $L_w$ as the relevant rod length scale governing the conductivity of polydispersed rod-like particles, to the best of our knowledge there are no experiments specifically designed to verify this prediction in real composites. In particular, although deviations from an ideally homogeneous dispersion of the nanotubes are common in CNT-polymer composites[2–4], their effect on the predicted $L_w$-scaling is currently unknown.

Here we demonstrate that conductivities measured in CNT-epoxy composites with different, purposely tailored length distributions of the CNTs, but with equal concentration of nanotubes, do not scale with $L_w$, but rather follow a quasi-universal dependence upon the number average of the CNT lengths, $L_n = \langle L \rangle$, regardless of the particular distribution function of $L$. By combining inter-particle tunnelling with a generalised connectedness percolation theory, we explain the observed $L_n$-scaling of $\sigma$ in terms of local clusters of tightly interlaced nanotubes present in our samples, whose effect is to change the relevant CNT length scale from $L_w$ to $L_n$. Our theoretical and experimental results suggest that the conductivity of CNT nanocomposites, if processed to enhance clustering of the nanotubes, can be made practically insensitive to the scaled variance of the CNT length distribution.

## Results

**CNT-epoxy composites with unimodal distribution of nanotube lengths.** We synthesized multi-walled CNTs by catalytic chemical vapour deposition as described in ref.[27] and in the Method section. We produced five different batches of CNTs with specific nanotube length distributions by cutting the as-grown CNTs by planetary ball milling. The use of the ball milling apparatus enabled the tailoring of the CNT lengths through specific combinations of milling times and rotational speeds[28]. The resulting CNT length distributions obtained from about 500 manually measured nanotube lengths from SEM micrographs, are shown in the histograms of Fig. 1 for five different combinations of the milling time and rotational speed. We have labeled the different batches as indicated in Fig. 1. To a good approximation, all distributions follow a lognormal distribution function, as observed in a previous report[28] (Supplementary Fig. 1). The $L_w$–$L_n$ plot of Fig. 2 shows that the weighted and number averages of the CNT lengths (filled squares) decrease gradually from $L_w = 5166$ nm and $L_n = 2277$ nm ("Long" CNTs) to $L_w = 913$ nm and $L_n = 590$ nm ("Short" CNTs), respectively, as the milling time and the rotational speed change from 30 min at 200 rpm to 6 h at 400 rpm. The scaled variance $\langle L^2 \rangle/\langle L \rangle^2 - 1 = L_w/L_n - 1$ is maximum for the "Long" CNTs ($\simeq 130$ %) and minimum for the "Long" CNTs ($\simeq 55$ %) (Supplementary Fig. 2).

For each batch of CNTs with a specific length distribution, we fabricated CNT-epoxy nanocomposites by dispersing the nanotubes in a SU8 matrix. SU8 is an epoxy-based UV-sensitive photoresist, particularly suited for thick-film applications, with a versatile patternability even if loaded with nanoparticle (Fig. 3a). In the present study, the CNT concentration was kept at $x = 0.6$ %wt with respect to the weight of the SU8 resin, which corresponds to a CNT volume fraction $\phi = x\rho SU8/(\rho CNT + x\rho SU8) = 0.28\%$, where $\rho SU8 = 0.998$ g/cm$^3$ and $\rho_{CNT} = 2.1$ g/cm$^3$ are the mass densities of SU8 and CNTs, respectively[29]. The processing temperature was low enough to prevent cross-linking of the epoxy. Examples of the morphology of the so-obtained CNT-SU8 composites are shown in the SEM and TEM images of Fig. 3b–e. From the analysis of TEM images of microtome slices of the composites, we have determined that the CNT diameters (denoted $D$) follow a lognormal distribution with first and second moments of $D$ given by $\langle D \rangle = 16$ nm and $\langle D^2 \rangle = 346$ nm (Supplementary Fig. 3). Apart from the modified CNT length distribution, the so-obtained composites are similar to the non-polymerized CNT-SU8 samples studied in ref.[29]. These have been shown to exhibit a conductivity behaviour as a function of the CNT volume fraction that was consistent with a tunnelling-dominated transport mechanism.

The conductivity data obtained from 4-point-probe measurements of the five sets of CNT-SU8 composites are shown in Fig. 4 (filled squares) as a function of the inverse of the weighted average $L_w$ of the CNT lengths. The conductivity gradually increases as $L_w$ is enhanced, which is consistent with the general trend expected in dispersions of conducting rod-like particles. As mentioned above, theories and simulations on systems of polydispersed rods predict that the tunnelling conductivity depends on the length distribution of the rods only through $L_w$. This is readily seen by adopting the critical distance approximation for $\sigma$[8,12,30–32]

$$\sigma \simeq \sigma_0 e^{-2\delta_c/\xi}, \qquad (1)$$

where $\sigma_0$ is a conductivity prefactor and $\xi$ is the localization length, and by using

$$\delta_c \simeq \frac{\langle D^2 \rangle}{2L_w \phi} \qquad (2)$$

for the critical tunnelling distance[11,13,26]. The latter expression derives from the second virial approximation of the liquid-state integral equation theory of percolation applied to a system of tunnelling connected rods homogeneously dispersed in an insulating matrix[18,22,23,25]. Equation 2 can also be derived from the Bethe lattice approach[19–21], or from the random geometric graph theory of ref.[35]. All these methods rely on the irrelevance of closed loops in the network formed by connected rods of asymptotically large aspect ratios[35].

Although equations (1) and (2) predict a linear decrease of $ln(\sigma)$ with $1/L_w$, the data of Fig. 4 follow a straight line only if we exclude the case with the shortest $L_w$. A linear fit to the remaining $ln(\sigma)$ data gives a slope equal to about $m = -9200$ nm (inset of Fig. 4). Using $\phi = 0.28\%$ and $\langle D^2 \rangle \simeq 346$ nm$^2$, this value of $m$ leads to $\xi \approx 13$ nm. While being 5 to 10 times larger than the expected value of the localization length for polymer composites, this value of $\xi$ is not large enough to invalidate the theory. Deviations from idealized models not considered in equation (2) may indeed account for this quantitative discrepancy[8].

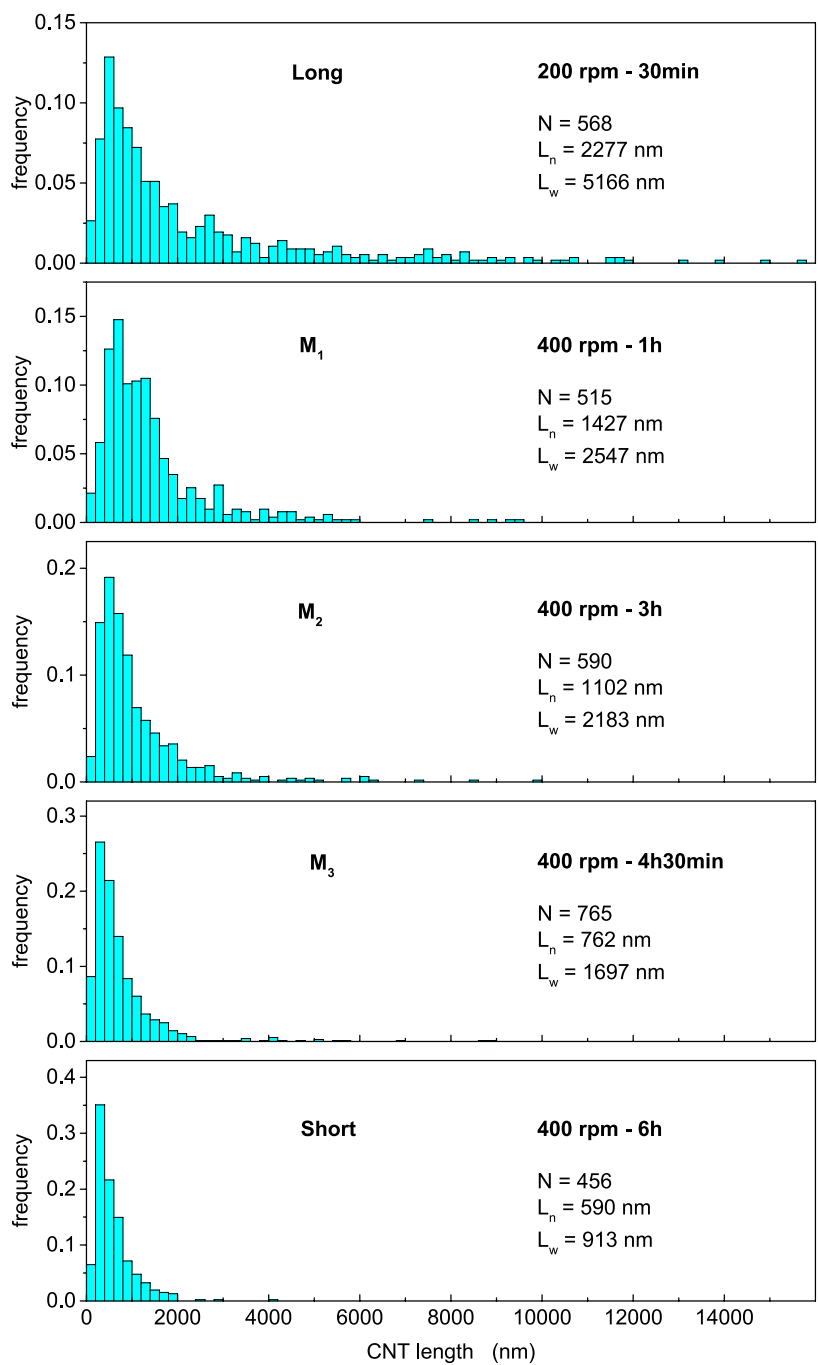

**Figure 1.** Normalized frequency of the as-milled CNT lengths as measured from SEM images. From the top panel to the bottom panel, the distributions become increasingly narrow.

**Bimodal distribution of CNT lengths.** To make a more stringent test, we fabricated additional CNT-SU8 composites with specifically engineered nanotube length dispersions. The rationale behind this approach is that the predicted scaling of $\sigma$ with $1/L_w$ should not depend on the specific distribution function of $L$ (which is approximately lognormal for the as-milled samples of Fig. 1) and that the same scaling should therefore manifest also for other types of the CNT length distribution. To verify this property, we have produced samples of CNT-SU8 by mixing nanotubes taken from the Long and Short batches of the as-milled CNTs with chosen values $p$ of the number fraction of the Long nanotubes. In this way, the distribution of the nanotube lengths becomes bimodal: $\rho_p(L) = p\rho_{\text{Long}}(L) + (1-p)\rho_{\text{Short}}(L)$, where $\rho_{\text{Long}}$ and $\rho_{\text{Short}}$ are the length distribution functions of the Long and Short CNT batches, respectively. Characterisation by SEM analysis shows that the so-obtained CNT length distribution follows the predicted bimodal distribution function (Supplementary Fig. 4). In preparing the new

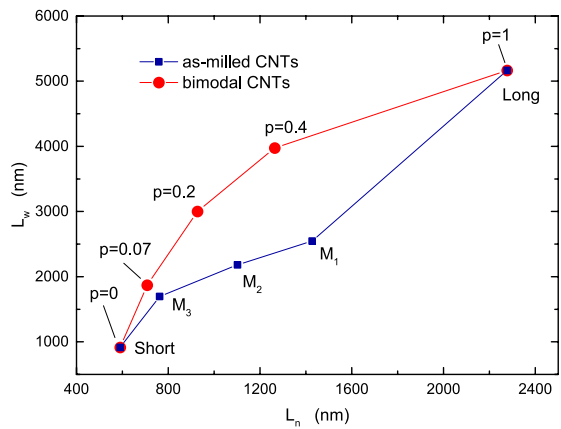

**Figure 2.** The weighted average $L_w = \langle L^2 \rangle / \langle L \rangle$ and the number average $L_n = \langle L \rangle$ of the CNT lengths for the as-milled nanotubes (filled squares) and for the bimodal length distributions (filled circles).

CNT-SU8 samples, we have taken the same CNT concentration of the as-milled composites ($x = 0.6$ %wt) with $p = 0.07$, $0.2$, and $0.4$. The resulting number and weighted length averages, obtained respectively from $L_n(p) = p\langle L\rangle_{\text{Long}} + (1-p)\langle L\rangle_{\text{Short}}$ and $L_w(p) = [p\langle L^2\rangle_{\text{Long}} + (1-p)\langle L^2\rangle_{\text{Short}}]/L_n(p)$ where the subscript "Long" ("Short") denotes an average over $\rho_{\text{Long}}(L)$ [$\rho_{\text{Short}}(L)$], are shown in Fig. 2 (filled circles). Varying the fraction of Long CNTs results in a trajectory in the $L_w$–$1/L_n$ plot (filled circles) that is different from that of the as-milled CNT batches, as seen in Fig. 2. Furthermore, the scaled variance of the bimodal CNTs is always larger than than of the unimodal nanotubes, attaining $\simeq 220$ % for $p = 0.2$ (Supplementary Fig. 2).

When plotted as a function of $1/L_w$, the conductivity behaviour of CNT-SU8 composites with bimodal CNT distributions differs from that of the CNT-SU8 samples with as-milled CNTs. This is clearly seen in Fig. 4 where $\sigma$ for the bimodal samples (filled circles) is well below that of the as-milled systems (filled squares) for comparable values of $1/L_w$. The finding that the conductivities of the two sets of composites fail to follow a common curve when plotted as a function of $1/L_w$ is in conflict with the prediction of model systems of polydispersed rods, and it implies that the agreement between theory and experiment discusses above is, actually, only apparent.

Surprisingly, although our CNT-SU8 composites do not follow the predicted $1/L_w$ scaling, they do however show a clear universal behaviour with respect to the inverse of the mean CNT length $L_n = \langle L \rangle$. This is demonstrated in Fig. 5a where we have replotted the measured values of $\sigma$ as a function of $1/L_n$: regardless of the specific distribution function of the CNT lengths, the conductivity data collapse into a single curve as a function of $1/L_n$.

**Modelling of the critical tunnelling distance.** Figure 5a implies that the rod length scale that is relevant for transport in CNT-SU8 composites is $L_n$ rather than $L_w$. To explain this result we model the critical tunnelling distance $\delta_c$ by relaxing the requirement, from which equation (2) is derived, that the nanotubes in our CNT-SU8 composites can be approximated as straight rod-like particles that are homogeneously dispersed in the epoxy. Deviations from such an idealized morphology are common in CNT-polymer nanocomposites, as they are very sensitive to the chemical composition of the matrix and to the processing history of the material[2,4,33,34]. In particular, SEM and TEM images of our CNT-SU8 composites (Fig. 3) show that the nanotubes form spaghetti-like networks with local clusters of tightly interlaced CNTs. The resulting network morphology is schematically represented in Fig. 5b which shows CNT clusters connected by dispersed nanotubes.

To model the electrical connectedness associated to the network topology of Fig. 5b, we make use of the random geometrical graph theory of continuum percolation[35] to account for the waviness of the nanotubes and the local CNT clusters. The latter are modelled by a phenomenological contact term in the pair distribution function of the nanotubes to account for the enhanced probability of finding nanotubes at contact when clusters are present. In this way, we obtain that the mean number of nanotubes of length $L'$ which are at distance $\delta$ from a given nanotube of length $L$ is (see Methods)

$$Z(L, L') = Z_{\text{cl}} + \frac{\pi}{2}\rho L L' \delta, \qquad (3)$$

where $\rho$ is the number density of the CNTs and $Z_{\text{cl}}$ is the mean coordination number of nanotubes at contact, which we treat for the moment as a constant. The large aspect ratio of the nanotubes justifies a tree-like approximation for the network, from which we obtain that the mean size of the component connected to the selected nanotube is

$$T(L) = 1 + \langle Z(L, L') T(L') \rangle', \qquad (4)$$

where $\langle \cdots \rangle'$ denotes an average over $L'$. The critical tunnelling distance $\delta_c$ is the smallest value of $\delta$ such that $\langle T(L) \rangle$ diverges. From equations (3) and (4) we obtain therefore the following analytical formula for $\delta_c$:

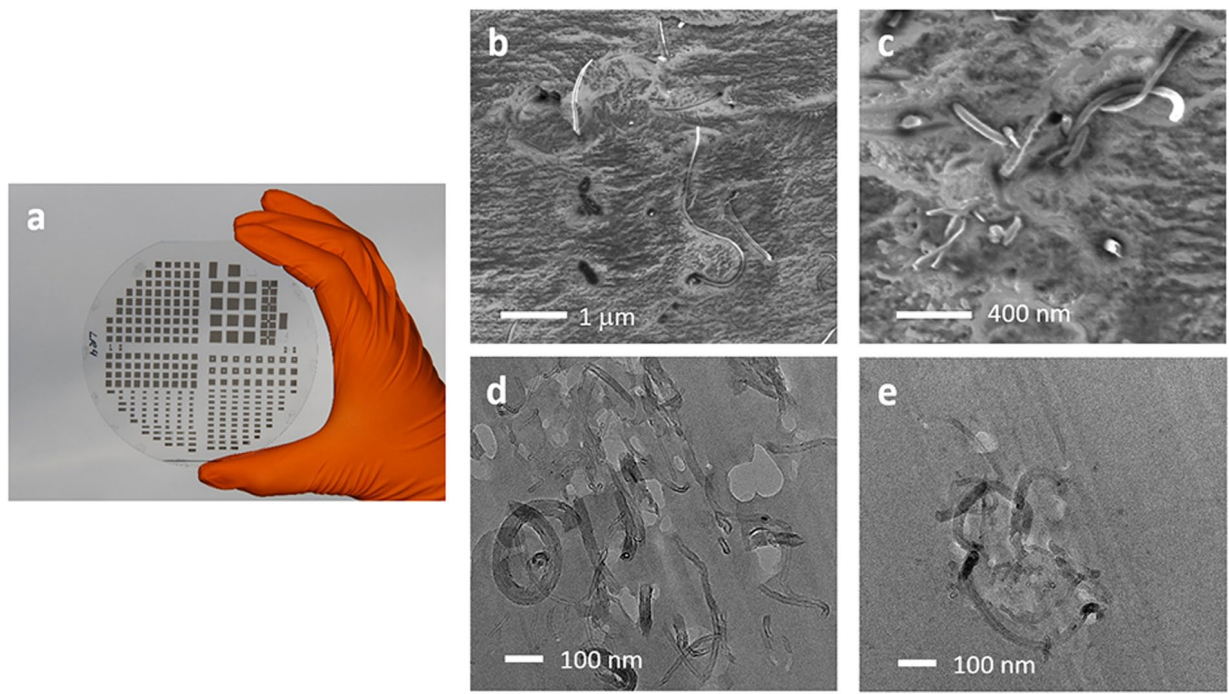

**Figure 3.** (**a**) SU8 patterned by UV-lithography process loaded with nanoparticles (0.6 wt% of reduced graphene oxide). (**b**,**c**) SEM images of two regions of the same CNT-SU8 sample. In (**b**) the nanotubes appear to be well dispersed while **c** shows a local cluster of tightly interlaced CNTs. (**d**,**e**) TEM images showing local CNT clusters. In (**b**–**e**) the CNT volume fraction is 0.28 %.

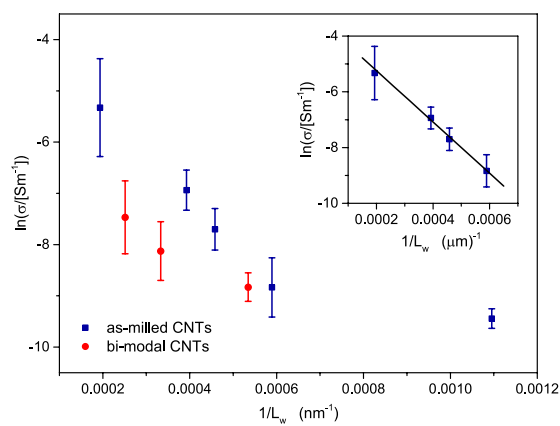

**Figure 4.** Natural logarithm of the conductivity measured in CNT-SU8 nanocomposites as a function of $1/L_w$. Each symbol represents an average over five independent conductivity measurements. Data for the composites with as-milled CNTs are shown by filled squares, while those with bi-modal CNT length distributions are shown by filled circles. Inset: linear fit of the as-milled samples.

$$\delta_c = \frac{\langle D^2 \rangle}{2\phi} \frac{1 - Z_{cl}}{(1 - Z_{cl})L_w + Z_{cl}L_n}, \quad (5)$$

where $1 - Z_{cl} > 0$ to ensure that $\delta_c$ is positive and $\phi = \rho(\pi/4)\langle D^2 \rangle \langle L \rangle$ is the volume fraction of the CNTs. Despite its simplicity, equation (5) is a non-trivial result showing the effect that clustering has on the CNT length scale relevant for percolation. The limit of ideally homogeneous distributions of CNTs (that is, no clustering) is obtained by setting $Z_{cl} = 0$ in equation (5), from which we recover equation (2). In this limit, therefore, the relevant length scale for transport at a given $\phi$ is $L_w$. Allowing for particle clustering ($Z_{cl} \neq 0$) changes qualitatively the role that nanotube polydispersity has on $\delta_c$. Remarkably, at the lowest order in $1 - Z_{cl}$ we find from equation (5) that:

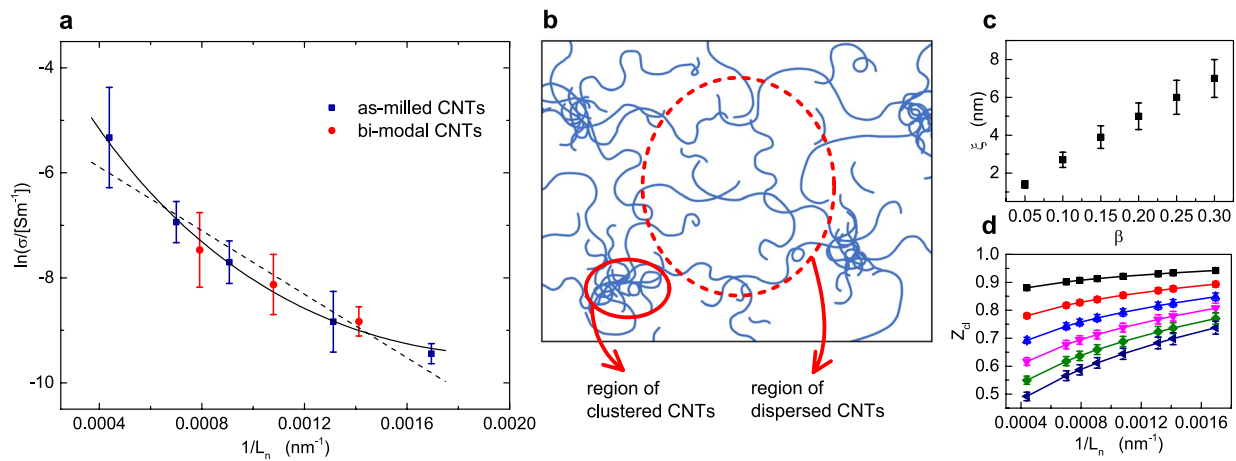

**Figure 5.** (**a**) Natural logarithm of the conductivity measured in CNT-SU8 nanocomposites as a function of $1/L_n$ for the as-milled (filled squares) and the bi-modal (filled circles) CNTs. Both sets of data follow a common curve, which evidences a quasiuniversal behaviour that is independent of the particular distribution of the CNT lengths. The dashed line is a linear fit to equation (6) where the clustering factor $Z_{cl}$ is assumed to be a constant. The solid line is a non-linear fit obtained by assuming that $Z_{cl}$ scales as $1/L_n^\beta$, with $\beta = 1/10$. (**b**) Schematic representation of the CNT network: the nanotubes form local clusters in the epoxy matrix. (**c**) Values of the localisation length $\xi$ obtained from nonlinear fits of the conductivity for different values of the exponent $\beta$. (**d**) Clustering factor as a function of $1/L_n$ for $\beta$ ranging from 0.05 to 0.3 (from top to bottom).

$$\delta_c = \frac{\langle D^2 \rangle}{2\phi L_n}(1 - Z_{cl}),$$
(6)

which predicts that $\delta_c$, and so the conductivity, is an universal function of $L_n = \langle L \rangle$ in this regime.

What equation (6) hints at is that the common dependence of $\sigma$ upon $L_n$ exhibited by both the as-milled and bimodal samples of CNT-SU8 composites originates from particle clustering effects. A linear fit of the data of Fig. 5a to $ln(\sigma)$ vs. $1/L_n$ (dashed line) gives a slope of about $m = -3500$ nm which, using equations (1) and (6), corresponds to a localization length of about $\xi = \frac{\langle D^2 \rangle}{\phi m}(1 - Z_{cl}) \simeq 35(1 - Z_{cl})$ nm. A realistic value of $\xi$ ($\xi \sim 3$–$4$ nm) can therefore be deduced by a clustering parameter of about $Z_{cl} = 0.9$, suggesting a rather important deviation from an ideal homogeneous dispersion of the CNTs in the SU8 epoxy.

The linear dependence of $ln(\sigma)$ upon $1/L_n$ predicted by equation (6) is only in partial agreement with the experimental data of Fig. 5a, which show a positive curvature as a function of $1/L_n$. The fit can be greatly improved by allowing a weak dependence of $Z_{cl}$ on the CNT lengths, which we parametrise by assuming that $Z_{cl}$ scales as $L_n^{-\beta}$. For $\beta = 1/10$, the best fit is obtained for $\xi \simeq 2.7$ nm (solid line in Fig. 5a) with the resulting value of $Z_{cl}$ being in the range 0.9–0.8 for all $L_n$. Equally accurate fits can be achieved by choosing different values of the exponent $\beta$, which gives the corresponding $\xi$ and $Z_{cl}$ shown in Fig. 5c,d. Provided that $\beta \lesssim 0.2$, we still obtain physically acceptable values of the localisation length (say, $\xi \lesssim 5$ nm) while $1 - Z_{cl} \lesssim 0.4$ in the entire range of $1/L_n$ probed by the experiments.

By using a different modelling of the clustering effect in a network of polydisperse rods, Chatterjee[26,36] has recently reached a result similar to the one presented here: namely, that enhanced clustering promotes a $L_n$-dependence of the electrical connectedness. In particular, the theory of refs[26,36] predicts that the critical distance, and so $ln(\sigma)$, is proportional to $1/L_n$ for sufficiently clustered rods, as in equation (6). There are however quantitative differences, due presumably to an approximation scheme different from the one employed here.

## Discussion

The size polydispersity of the nanotubes and their dispersion in the matrix are factors that strongly influence the conductivity characteristics of CNT-polymer nanocomposites. Our experiments on polydisperse CNT-SU8 materials and the theoretical modelling hint at a simple, yet comprehensive, understanding of these factors and of the role they have in the conductivity behaviour of the composite. Of particular interest for the optimisation of the transport properties of CNT-polymer materials is the interplay between particle size polydispersity and clustering that is exhibited by equation (5). Indeed, if on the one hand a pronounced CNT polydispersity in homogeneous dispersions generally enhances the conductivity because $L_w > L_n$, on the other hand clustering effects tend to marginalise the role of polydispersity by at the same time enhancing $\sigma$. It is thus possible to ignore the effect of particle polydispersity and have high levels of the conductivity if the CNT dispersion is sufficiently clustered. This has practical implications, because it allows to characterize clustered CNT-polymer composites through a single moment, $L_n$, of the nanotube lengths.

Since the CNT volume fraction in our composites was held constant, we did not study its effect on the length distribution dependence of $\sigma$. Depending on whether $Z_{cl}$ is affected or not by the CNT loading, we can expect two broad scenarios. In the first, the probability that a CNT belongs to a cluster diminishes as the CNT concentration is reduced. In this case, $Z_{cl}$ is roughly proportional to $\phi$ at low CNT loading, and our theory predicts that the conductivity eventually becomes dependent only on $L_w$ for values of $\phi$ sufficiently small such that $Z_{cl} \ll 1$. In the second scenario, the cluster number $Z_{cl}$ remains finite even as the volume fraction becomes arbitrarily small, as the nanoparticles may clump together quite strongly due to Van der Waals forces and/or covalent bonding. In this case, the conductivity may display a $1/L_n$-scaling regardless of the CNT volume fraction.

The ultimate proof of our claim would require an extensive experimental study involving the control of the dispersion of the nanotubes in the matrix. According to equation (5), indeed, the relevant CNT length scale would change from $\sim L_n$ to $\sim L_w$ as the particle dispersion in the matrix could be varied from cluster-dominated to homogeneous. Finally, we point out that since local clustering of the nanotubes is typical in real materials[34], the combined effect of polydispersity and clustering discussed here should be a common feature of other CNT-polymer nanocomposites, which could be tuned by the morphological control of the conductive network[4].

## Methods

### CNT synthesis and ball milling.
Multiwalled CNTs were synthesized by catalytic chemical vapour deposition of acetylene using Fe-Co catalytic particles supported by calcite. The synthesis of CNTs was carried out in a horizontally mounted quartz furnace at 720 °C under flow of acetylene and Nitrogen for 2 hours. In order to remove the catalytic particles and the supporting material, as-grown CNTs were purified by stirring in hydrochloric acid of 1 Molar, filtered and washed with distilled water and ethanol. The average length and diameter of the as-produced CNTs were approximately 10 m and 16 nm, respectively.

The as-grown CNTs were cut by planetary ball milling in a liquid environment (Gamma-butyrolactone). We have processed batches consisting of $ZrO_2$ balls of 3 mm of radius, Gamma-butyrolactone, and CNTs in mass proportion of 40:20:1 in 250 ml zirconium oxide-lined jars.

### Preparation of the CNT-SU8 composites.
CNTs of given length distributions were dispersed by sonication in the presence of surfactant in a SU8 epoxy matrix (Gersteltec, grade GM1060), which is constituted by SU8 resin and 40% of solvent. Subsequently, the ink was spread on a clean glass slide by doctor blading and was soft-baked following a temperature ramp lasting 15 min up to 95 °C to evaporate the solvent.

### CNTs with bimodal distribution of lengths.
We produced CNTs with bimodal distributions of the nanotube lengths by mixing the Long and Short batches of the as-milled CNTs with chosen values of the number fraction $p = N_{Long}/(N_{Long} + N_{Short})$, where $N_{Long}$ and $N_{Short}$ are the number of the Long and Short nanotubes, respectively. In practice, for a given value of $p$ we took the weight ratio of the Long nanotubes over the Short nanotubes such that

$$\frac{N_{Long}}{M_{Short}} = \frac{N_{Long}\langle L \rangle_{Long}}{N_{Short}\langle L \rangle_{Short}} = \frac{p}{1-p}\frac{\langle L \rangle_{Long}}{\langle L \rangle_{Short}},$$

where $\langle L \rangle_{Long} \sim 2277$ nm and $\langle L \rangle_{Short} \sim 590$ nm are the mean nanotube lengths measured from the Long and Short CNT batches, respectively.

### Calculation of the critical distance.
Two nanotubes are considered as connected if the separation between their closest surfaces is smaller than a given distance $\delta$. The critical distance $\delta_c$ is defined as the minimal distance such that a giant component of connected nanotubes exists. We model the nanotubes as wormlike impenetrable cylinders with distributed lengths and diameters and calculate $\delta_c$ by using the random geometric graph theory of continuum percolation[35,37]. The advantage of this method compared to the integral equation theory is that there is no need to invoke thermodynamic equilibrium of the nanoparticles. Let us introduce the probability $p_{ij}$ that a cylinder of length $L_j$ and diameter $D_j$ is found within a distance $\delta$ from a given cylinder of length $L_i$ and diameter $D_i$. Using a curvilinear coordinate system[38], the probability can be written as:

$$p_{ij} = \frac{2}{V}\int_0^{L_i} ds \int_0^{L_j} ds' \int_{D_{ij}}^{D_{ij}+\delta} d\Delta \langle |sin\gamma(\hat{u}(s), \hat{u}(s'))| \\ \times g_{ij}^{(2)}(\Delta, \hat{u}(s), \hat{u}(s'))\rangle, \quad (7)$$

where $V$ is the volume of system, $\Delta$ is the distance between the closest contour points $s$ and $s'$, $D_{ij} = (D_i + D_j)/2$, $\hat{u}(s)$ and $\hat{u}(s')$ are unit tangent vector at the contour points, and $\gamma(\hat{u}(s), \hat{u}(s'))$ is the angle formed by $\hat{u}(s)$ and $\hat{u}(s')$. In equation (7) $g_{ij}^{(2)}$ denotes the pair distribution function which is proportional to the probability of finding two cylinders at relative distance $\Delta$ and local orientations $\hat{u}(s)$ and $\hat{u}(s')$, and $\langle \cdots \rangle$ is an average with respect to the waviness of the cylinders. The mean number of nanotubes at distance $\delta$ from the given nanotube is $Z_{ij} = Nx_j p_{ij}$, where $N$ is the total number nanotubes and $x_j$ is the fraction of nanotubes of length $L_j$ and diameter $D_j$.

To account for the existence of local clusters of nanotubes we use a minimal model in which $g_{ij}^{(2)}$ is strongly peaked at $\Delta = D_{ij}$ and $g_{ij}^{(2)} \simeq 1$ for $\Delta > D_{ij}$. In this way $Z_{ij}$ can be written as: $Z_{ij} = x_j Z(L_i, L_j)$, where $Z(L_i, L_j) = Z_{cl} + \rho \pi L_i L_j \delta/2$, where $\rho = N/V$, $Z_{cl}$ is the number of cylinders belonging to a cluster, and where we have set $\langle sin\gamma \rangle \simeq \pi/4$ in equation (7) because the average with respect to the waviness of the cylinders is essentially an average over an isotropic orientation of the tangent vectors $\hat{u}(s)$ and $\hat{u}(s')$.

If we neglect the contribution of closed loops of connected cylinders, then the mean size $S_i$ of the component to which the selected cylinder belongs is $S_i = x_i + x_i \sum_j Z_{ij} T_j$, where

$$T_j = 1 + \sum_l Z_{jl} T_l, \qquad (8)$$

is the mean size of the branches connected to the the $j$-th cylinder. Equation (8) can be rewritten as in equation (4) if we define the average over the nanotube lengths as $\langle (\cdots) \rangle = \sum_i x_i (\cdots)$. The critical distance $\delta_c$ is obtained by requiring that $S = \sum_i S_i$ diverges, which, using equation (8), is equivalent to ask that $S = \sum_i x_i T_i = \langle T(L) \rangle$ goes to infinity.

## References

1. Coleman, J. N., Khan, U., Blau, W. J. & Gun'ko, Y. K. Small but strong: a review of the mechanical properties of carbon nanotube-polymer composites. *Carbon* **44**, 1624–1652 (2006).
2. Bauhofer, W. & Kovacs, J. Z. A review and analysis of electrical percolation in carbon nanotube polymer composites. *Compos. Sci. Technol.* **69**, 1486–1498 (2009).
3. Byrne, M. T. & Gun'ko, Y. K. Recent advances in research on carbon nanotube-polymer composites. *Adv. Mater.* **22**, 1672–1688 (2010).
4. Deng, H. *et al.* Progress on the morphological control of conductive network in conductive polymer composites and the use as electroactive multifunctional materials. *Prog. Polym. Sci.* **4**, 627–655 (2014).
5. Mutiso, R. M. & Winey, K. I. Electrical Properties of Polymer Nanocomposites Containing Rod-Like Nanofillers. *Prog. Polym. Sci.* **40**, 63–84 (2015).
6. Balberg, I., Anderson, C. H., Alexander, S. & Wagner, N. Excluded volume and its relation to the onset of percolation. *Phys. Rev. B* **30**, 3933–3943 (1984).
7. Bug, A. L. R., Safran, S. A. & Webman, I. Continuum Percolation of Rods. *Phys. Rev. Lett.* **54**, 1412–1315 (1985).
8. Ambrosetti, G. *et al.* Solution of the tunneling-percolation problem in the nanocomposite regime. *Phys. Rev. B* **81**, 155434 (2010).
9. Safdari, M. & Al-Haik, M. Electrical conductivity of synergistically hybridized nanocomposites based on graphite nanoplatelets and carbon nanotubes. *Nanotechnol.* **23**, 405202 (2012).
10. Alvarez, C. E. & Klapp, S. H. Percolation and orientational ordering in systems of magnetic nanorods. *Soft Matter* **8**, 7480–7489 (2012).
11. Nigro, B., Grimaldi, C., Ryser, P., Chatterjee, A. P. & van der Schoot, P. Quasiuniversal connectedness percolation of polydisperse rod systems. *Phys. Rev. Lett.* **110**, 015701 (2013).
12. Nigro, B., Grimaldi, C., Miller, M. A., Ryser, P. & Schilling, T. Depletion-interaction effects on the tunneling conductivity of nanorod suspensions. *Phys. Rev. E* **88**, 042140 (2013).
13. Kale, S., Sabet, F. A., Jasiuk, I. & Ostoja-Starzewski, M. Tunneling-percolation behavior of polydisperse prolate and oblate ellipsoids. *J. Appl. Phys.* **118**, 154306 (2015).
14. Kale, S., Sabet, F. A., Jasiuk, I. & Ostoja-Starzewski, M. Effect of filler alignment on percolation in polymer nanocomposites using tunneling-percolation model. *J. Appl. Phys.* **120**, 045105 (2016).
15. Inam, F., Reece, M. J. & Peijs, T. Shortened carbon nanotubes and their influence on the electrical properties of polymer nanocomposites. *J. Compos. Mater.* **46**, 1313–1322 (2012).
16. Castillo, F. Y. *et al.* Electrical, mechanical, and glass transition behavior of polycarbonate-based nanocomposites with different multi-walled carbon nanotubes. *Polymer* **52**, 3835–3845 (2011).
17. Deng, H., Zhang, R., Bilotti, E., Loos, J. & Peijs, T. Conductive polymer tape containing highly oriented carbon nanofillers. *J. Appl. Polym. Sci.* **113**, 742–751 (2009).
18. Kyrylyuk, A. V. & van der Schoot, P. Continuum percolation of carbon nanotubes in polymeric and colloidal media. *Proc. Natl. Acad. Sci. USA* **105**, 8221–8226 (2008).
19. Chatterjee, A. P. Percolation thresholds for rod-like particles: polydispersity effects. *J. Phys.: Condens. Matter* **20**, 255250 (2008).
20. Chatterjee, A. P. Connectedness percolation in polydisperse rod systems: A modified Bethe lattice approach. *J. Chem. Phys.* **132**, 224905 (2010).
21. Chatterjee, A. P. A Remark Concerning Percolation Thresholds in Polydisperse Systems of Finite-Diameter Rods. *J. Stat. Phys.* **146**, 244–248 (2012).
22. Otten, R. H. J. & van der Schoot, P. Continuum Percolation of Polydisperse Nanofillers. *Phys. Rev. Lett.* **103**, 225704 (2009).
23. Otten, R. H. J. & van der Schoot, P. Connectivity percolation of polydisperse anisotropic nanofillers. *J. Chem. Phys.* **134**, 094902 (2011).
24. Mutiso, R. M., Sherrott, M. C., Li, J. & Winey, K. I. Simulations and generalized model of the effect of filler size dispersity on electrical percolation in rod networks. *Phys. Rev. B* **86**, 214306 (2012).
25. Meyer, H., van der Schoot, P. & Schilling, T. Percolation in suspensions of polydisperse hard rods: Quasi universality and finite-size effects. *J. Chem. Phys.* **143**, 044901 (2015).
26. Chatterjee, A. P. A percolation-based model for the conductivity of nanofiber composites. *J. Chem. Phys.* **139**, 224904 (2013).
27. Smajda, R., Mionić, M., Duchamp, M., Andresen, J. C., Forró, L. & Magrez, A. Production of high quality carbon nanotubes for less than $1 per gram. *Phys. Status Solidi (c)* **7**, 1236–1240 (2010).
28. Forró, L., Gaal, R., Grimaldi, C. & Mionić, M. Tuning the length dispersion of multi-walled carbon nanotubes by ball milling. *AIP Adv.* **3**, 092117 (2013).
29. Grimaldi, C., Mionić, M., Gaal, R., Forró, L. & Magrez, A. Electrical conductivity of multi-walled carbon nanotubes-SU8 epoxy composites. *Appl. Phys. Lett.* **102**, 223114 (2013).
30. Nigro, B., Ambrosetti, G., Grimaldi, C., Maeder, T. & Ryser, P. Transport properties of nonhomogeneous segregated composites. *Phys. Rev. B* **83**, 064203 (2011).
31. Nigro, B., Grimaldi, C., Miller, M. A., Ryser, P. & Schilling, T. Tunneling conductivity in composites of attractive colloids. *J. Chem. Phys.* **136**, 164903 (2012).
32. Nigro, B. *et al.* Enhanced tunneling conductivity induced by gelation of attractive colloids. *Phys. Rev. E* **87**, 062312 (2013).
33. Alig, I. *et al.* Establishment, morphology and properties of carbon nanotube networks in polymer melts. *Polymer* **53**, 4–28 (2012).
34. Gnanasekaran, K., de With, G. & Friedrich, H. Quantitative Analysis of Connectivity and Conductivity in Mesoscale Multiwalled Carbon Nanotube Networks in Polymer Composites. *J. Phys. Chem. C* **120**, 27618–27627 (2016).
35. Chatterjee, A. P. & Grimaldi, C. Random geometric graph description of connectedness percolation in rod systems. *Phys. Rev. E* **92**, 032121 (2015).
36. Chatterjee, A. P. Geometric percolation in polydisperse systems of finite-diameter rods: Effects due to particle clustering and inter-particle correlations. *J. Chem. Phys.* **137**, 134903 (2012).
37. Grimaldi, C. Continuum percolation of polydisperse hyperspheres in infinite dimensions. *Phys. Rev. E* **92**, 012126 (2015).
38. Jinbo, Y., Sato, T. & Teramoto, A. Light Scattering Study of Semiflexible Polymer Solutions. 1. Dilute through Semidilute Solutions of Poly(n-hexyl isocyanate) Dissolved in Dichloromethane. *Macromolecules* **27**, 6080–6087 (1994).

### Acknowledgements
The authors thank Avik P. Chatterjee and P. van der Schoot for useful comments, R. Gaal for assistance in the conductivity measurements, E. Kecsnovity for assistance in the synthesis of CNTs, and J. F. G. Marin for assistance in the ball milling and the SEM image analysis. The Swiss National Science Foundation supported this work (Grant No. 200021-140557).

### Author Contributions
M.M. synthesised the materials, characterised the CNTs, and performed the experiments, A.M. coordinated the experiments. C.G. suggested the experiment and formulated the theory. L.F. coordinated the research. All authors discussed the results and contributed to the writing of the manuscript.

### Additional Information
**Supplementary information** accompanies this paper at https://doi.org/10.1038/s41598-017-12857-8.

**Competing Interests:** The authors declare that they have no competing interests.

**Publisher's note:** Springer Nature remains neutral with regard to jurisdictional claims in published maps and institutional affiliations.